\documentclass[runningheads]{llncs}
\usepackage{graphicx}
\usepackage{algorithm}
\usepackage{algpseudocode}
\usepackage{amsmath}
\usepackage{bbm}
\usepackage{latexsym,amsfonts,amssymb}
\usepackage{threeparttable}
\usepackage{array}

\newcolumntype{P}[1]{>{\centering\arraybackslash}p{#1}}

\begin{document}
\title{Revisiting Rubik's Cube: Self-supervised Learning with Volume-wise Transformation for 3D Medical Image Segmentation}
\titlerunning{ }

\author{Xing Tao\inst{1}\thanks{This work was done when Xing Tao was an intern at Tencent Jarvis Lab} \and Yuexiang Li\inst{2} \and Wenhui Zhou\inst{1} \and Kai Ma\inst{2} \and Yefeng Zheng\inst{2}}
\authorrunning{X. Tao et al.}
\institute{School of Computer Science and Technology, Hangzhou Dianzi University, Hangzhou, China \\
\email{zhouwenhui@hdu.edu.cn} \and
Tencent Jarvis Lab, Shenzhen, China\\
\email{vicyxli@tencent.com}
}

%
%
%
\maketitle              
\begin{abstract}
  Deep learning highly relies on the quantity of annotated data. However, the annotations for 3D volumetric medical data require experienced physicians to spend hours or even days for investigation. Self-supervised learning is a potential solution to get rid of the strong requirement of training data by deeply exploiting raw data information. In this paper, we propose a novel self-supervised learning framework for volumetric medical images. Specifically, we propose a context restoration task, i.e., Rubik's cube++, to pre-train 3D neural networks. Different from the existing context-restoration-based approaches, we adopt a volume-wise transformation for context permutation, which encourages network to better exploit the inherent 3D anatomical information of organs. Compared to the strategy of training from scratch, fine-tuning from the Rubik's cube++ pre-trained weight can achieve better performance in various tasks such as pancreas segmentation and brain tissue segmentation. The experimental results show that our self-supervised learning method can significantly improve the accuracy of 3D deep learning networks on volumetric medical datasets without the use of extra data.

\end{abstract}
\section{Introduction}

The success of convolutional neural networks (CNNs) benefits from the amount of annotated data rapidly increased in the last decade. However, the high-quality medical image annotations are extremely laborious and usually hard to acquire, which require hours or even days for an experienced physician. As a result, the limited quantity of annotated medical images is the major obstacle impeding the improvement of diagnosis accuracy with the latest 3D CNN architectures~\cite{CicekO01,MilletariF01}.

To deal with the problem of deficient annotated data, self-supervised learning approaches, which utilize unlabelled data to train network models in a supervised manner, attract lots of attentions. The typical self-supervised learning defines a relevant pretext task to extract meaningful features from unlabelled data, where the learned representations can boost the accuracy of the subsequent target task with limited training data. Various pretext tasks have been proposed, including patch relative position prediction~\cite{Doersch_2015_ICCV} (Jigsaw puzzles~\cite{NorooziVFP18} can be grouped into this category), grayscale image colorization~\cite{larsson_colorization_2017}, and context restoration~\cite{pathakCVPR16context}. The idea of self-supervised learning was firstly brought to medical image analysis by Zhang et al.~\cite{Zhang_2017}. They pre-trained a 2D network for fine-grained body part recognition with a pretext task that sorted the 2D slices from the conventional medical volumes. Compared to 2D networks, 3D networks integrating more spatial context information have shown the superiority for the 3D medical data~\cite{dou20173d}.

Recently, several studies have made their efforts to develop self-supervised learning frameworks for 3D neural networks~\cite{Spitzer_2018,Zhou_2019_MICCAI}. Zhuang et al.~\cite{Zhuang_2019_MICCAI} proposed a pre-train 3D networks by playing a Rubik's cube game, which can be seen as an extension of 2D Jigsaw puzzles~\cite{Noroozi_2016_ECCV}. Formulated as a classification task, however, their model only pre-trained the down-sampling layers of CNNs. When applying the pre-trained weights to a target task requiring up-sampling operations (e.g., organ segmentation), the performance improvement is neutralized by the randomly initialized up-sampling layers. To this end, we reformulate the Rubik's cube game as a context restoration task, which simultaneously pre-trains the down-sampling and up-sampling layers of fully convolutional networks (FCNs).

In this paper, we propose a novel self-supervised learning pretext task, namely Rubik's cube++\footnote{The symbol ``++'' represents two improvements compared to the existing Rubik's cube~\cite{Zhuang_2019_MICCAI}: 1) encoder-decoder architecture, and 2) volume-wise transformation.}, for volumetric medical image segmentation. Inspired by the observation that learning from a harder task often leads to a more robust feature representation~\cite{Deng_2010_ECCV,Wei_2019_CVPR}, our Rubik's cube++ adopts a volume-wise transformation, e.g., 3D voxel rotation, to permute and restore the volumetric medical data, which is assumed to be much harder than the existing methods~\cite{Zhou_2019_MICCAI}. To validate our assumption that the voxel rotation is a better transformation for self-supervised learning with 3D data, we evaluate the proposed Rubik's cube++ on two medical image segmentation tasks (i.e., pancreas segmentation and brain tissue segmentation) using publicly available datasets. The experimental results demonstrate that our method can significantly boost the segmentation accuracy and achieve the state-of-the-art performance compared to other self-supervised baseline methods.

\begin{figure}[!t]
  \begin{center}
    \includegraphics[width=0.9\linewidth]{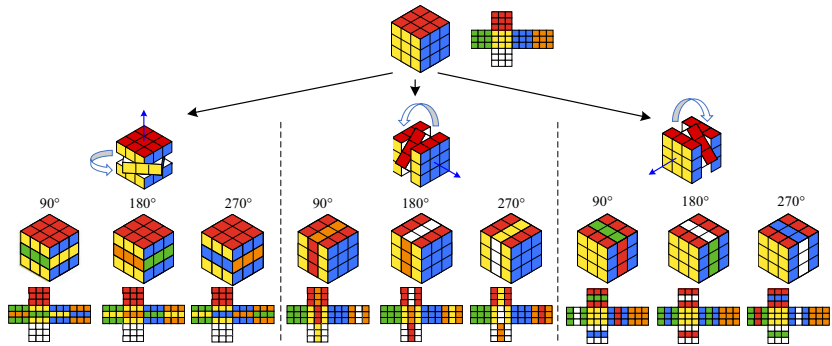}
  \end{center}\caption{The schematic diagram of the Rubik's cube++ transformations. The three-order Rubik's cube is taken as an example. There are three axes ($sagittal$, $coronal$ and $axial$) and three degree ($90^{\circ}$, $180^{\circ}$ and $270^{\circ}$) for each transformation to choose. The diversely transformed results are presented on the right, which prevent trivial solutions.}
  \label{fig:transform}
\end{figure}

\section{Method}
The volumetric medical data can be seen as a high-order Rubik's cube by partitioning it into a grid of subcubes. Let the dimension of the volumetric data be $W{\times}H{\times}L$ and the side length of the subcube be $n$ voxels, we have a Rubik's cube with $\lfloor\frac{W}{n}\rfloor \times \lfloor\frac{H}{n}\rfloor \times \lfloor\frac{L}{n}\rfloor$ subcubes, where $\lfloor\cdot\rfloor$ is the floor function. The initial state of these subcubes, including the order and orientation, is defined as the original state of Rubik's cube. We first clarify the definition of two components: subcube and cube layer.

\subsubsection{Subcube.} It is the smallest component containing 3D anatomical information. Different from \cite{Zhuang_2019_MICCAI}, the subcubes of our Rubik's cube are bound to their neighbors, which prohibits individual movement or rotation of the subcube.

\subsubsection{Cube layer.} As shown in Fig.~\ref{fig:transform}, the cube layer, consisting of a set of subcubes on the same anatomical plane along the major axies ($sagittal$, $coronal$ and $axial$), is used as the unit component for the transformation of Rubik's cube++.

Our Rubik's cube++ has a set of transformations. Take the $3\times3\times3$ Rubik's cube as an example, as shown in Fig.~\ref{fig:transform}, the cube layer containing nine subcubes can be rotated along a specific axis by a fixed angle. Note that if the Rubik's cube is of a cuboid shape, only $180^{\circ}$ rotation is valid along the short axis. Compared with the image transformations used in other pretext tasks \cite{Zhou_2019_MICCAI}, the transformations of Rubik's cube++ are restricted to volume-wise rotation. Such transformations wreck the 3D information of medical data, and encourage the network to exploit useful volumetric features for restoration. Moreover, the rotation operation can generate diversely transformed results without generating image artifacts, which prevents the network from learning trivial solutions \cite{Doersch_2015_ICCV}.

\subsection{Pretext Task: Rubik's Cube Restoration}
Our restoration-based pretext task is formulated to pre-train a 3D network. At the beginning of the pretext task, the Rubik's cube is permuted by a sequence of random volume-wise transformations, which results in a disarranged Rubik's cube. The original state of Rubik's cube is used as the supervision signal. A 3D network is trained to recover the original state of Rubik's cube from the disarranged state, which enforces the network to learn 3D anatomical information.

\begin{algorithm}[!tb]
  \caption{Disarrangement of Rubik's cube.}
  \label{alg:trans}
  \begin{algorithmic}[1]
    \Require
    Original 3D medical data ${\rm {\bf y}}$ of shape $W \times H \times L$.
    \State According to the preset parameter $n$, we partition the 3D
    medical data into a $\lfloor\frac{W}{n}\rfloor \times \lfloor\frac{H}{n}\rfloor \times \lfloor\frac{L}{n}\rfloor$ subcubes.
    \For{axis $i\in \{sagittal, coronal, axial\}$}
    \State Randomly designate $m$ cube layers along axis $i$.
    \For{layer $j\in \{1, 2,..,m\}$}
    \State Randomly select an angle $\theta\in \{90^{\circ},180^{\circ},270^{\circ}\}$.
    \State Rotate the cube layer $j$ along axis $i$ by angle $\theta$.
    \EndFor
    \EndFor
    \Ensure
    Disarranged 3D medical data ${\rm {\bf x}}$.
  \end{algorithmic}
\end{algorithm}

The process of Rubik's cube disarrangement adopted in Rubik's cube++ is presented in Algorithm~\ref{alg:trans}. Let $\bf y$ denote the 3D medical data, i.e., the original state of Rubik's cube. And, ${x}=T(y)$ indicates the 3D medical data in a disarranged state after a sequence of random transformations $T(\cdot)$. Note that we can easily adjust the difficulty of the pretext task by changing the parameters side length of subcube ($n$) and the number of rotated cube layers ($m$).

\subsection{Network Architecture}
\begin{figure*}[!tb]
  \begin{center}
    \includegraphics[width=0.95\linewidth]{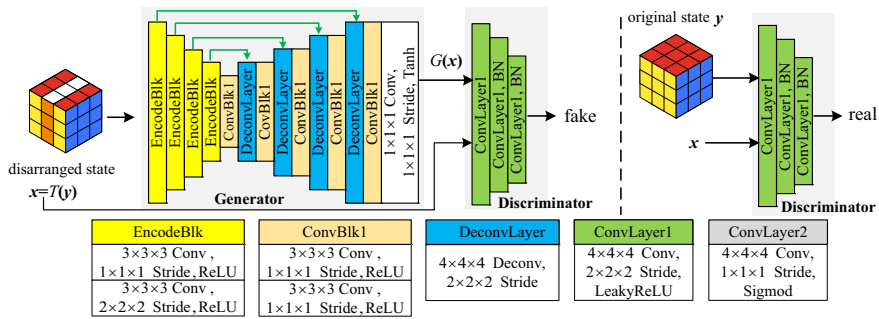}
  \end{center}
  \caption{The network architecture for our self-supervised learning.}
  \label{fig:network}
\end{figure*}

A GAN-based architecture is used to resolve our Rubik's cube++ pretext task, which consists of two components: a generator $G$ and a discriminator $D$. As shown in Fig.~\ref{fig:network}, both the generator and discriminator are of 3D neural networks. The generator adopts a 3D encoder-decoder structure with skip connections between mirrored layers in the encoder and decoder stacks, which is the same to \cite{CicekO01}. Note that other widely used 3D FCNs, such as V-Net \cite{MilletariF01}, can be easily adopted as the generator of our Rubik's cube++ for pre-training. The discriminator consists of four convolutional layers with the kernel size of $4\times 4\times 4$. The restored Rubik's cube $G(x)$ and original state $y$ are respectively concatenated with the disarranged state $x$ and fed to the discriminator for the real/fake classification.

The GAN-based framework aims to recover the original state of Rubik's cube from the disarranged state. We propose a joint loss function to supervise the Rubik's cube restoration. It consists of a reconstruction loss and an adversarial loss, which are responsible for capturing the overall structure of 3D medical data and tuning the anatomical details, respectively.

\paragraph{\bf Reconstruction loss.} As shown in Fig.~\ref{fig:network}, the disarranged state $x$ of Rubik's cube is fed to the generator for context restoration $G(x)$. The voxel-wise $\mathcal{L}_1$ loss between $y$ and $G(x)$ is calculated to optimize the restoration quality. We also tried to use $\mathcal{L}_{2}$ as the reconstruction loss for our Rubik's cube++.\footnote{An ablation study of $\mathcal{L}_1$ and $\mathcal{L}_2$ can be found in {\itshape arxiv version}} Compared to $\mathcal{L}_{1}$, the $\mathcal{L}_{2}$ loss is inclined to have blurry solutions, which may lose the boundary information of organs.

\begin{equation}
  \label{eq:l1}
  \mathcal{L}_{1} \left( G \right)={\mathbb E}_{{\rm {\bf x}},
  {\rm {\bf y}}}{\left\| {{\rm {\bf y}}-G\left( {\rm {\bf x}} \right)}
  \right\|_1}.
\end{equation}

\paragraph{\bf Adversarial loss.} The adversarial loss derived from \cite{Isola_2017_CVPR} is adopted in our pretext task to generate more real and elegant reconstruction results.\footnote{The reconstruction results are visualized in the {\itshape arxiv version}.} The generator learns a mapping from the disarranged state to the original state, $G:x \rightarrow y$. The discriminator learns to classify the fake (restored Rubik's cube produced by the generator) and the real (the original data). These two components are encouraged to compete against each other. The adversarial loss is defined as:
\begin{equation}
  \label{eq:lgan}
  \mathcal{L}_{adv}\left( {G,D} \right) =
  \mathbb{E}_{x,y}\log D\left( {x, y} \right)+
  \mathbb{E}_{x}\log \left( {1-D\left({{{x}},
      G\left({{x}} \right)}
    \right)} \right).
\end{equation}

\paragraph{\bf Objective.} For training, the generator is encouraged to fool the discriminator, while the discriminator is required to correctly classify the real and fake data. Therefore, $G$ tries to minimize $\mathcal{L}_{adv}$ and $\mathcal{L}_{1}$, while $D$ aims to maximize $\mathcal{L}_{adv}$. The full objective for our Rubik's cube++ restoration is summarized as:
\begin{equation}
  \label{eq:gloss1}
  \mathcal{L}=\arg \mathop {\min}\limits_G \mathop {\max}\limits_D
  \left( \mathcal{L}_{adv}\left( {G,D} \right)+\lambda {\mathcal L}_{1} \left( G \right) \right),
\end{equation}
where $\lambda$ is a tuneable hyperparameter which is set to 10 in our experiments.

\paragraph{\bf Transfer learning.} After the framework completes the self-supervised learning in the Rubik's cube++ restoration task, the trained generator that learns useful 3D anatomical information from raw data is adapted to the target task by replacing the last layer with the segmentation output layer. With the voxel-wise annotation, the pre-trained generator can be directly fine-tuned on the target segmentation task, which alleviates the influence caused by the randomly-initialized decoder \cite{Wei_2019_CVPR,Zhuang_2019_MICCAI}.

\section{Experiments}
\paragraph{\bf Datasets.} To evaluate the performance of our Rubik's cube++, we conduct a 4-fold cross validation on the NIH Pancreas computed tomography (CT) dataset~\cite{Roth_miccai_2015}, adopting the same protocol to \cite{Wei_2019_CVPR}. Moreover, a leave-one-out cross validation is conducted on the MRBrainS18 dataset~\cite{MRBrainS18}, due to its relatively small dataset size (i.e., only seven sets of brain magnetic resonance (MR) volumes). The 3D U-Net \cite{CicekO01} is used as backbone for the generator. The Dice coefficient (DSC)~\cite{Wei_2019_CVPR} is used to assess the segmentation accuracy. For the multi-class task (e.g., brain tissue segmentation on MRBrainS18), we calculate the DSC for each class and average them to yield a mean DSC for performance evaluation.

\paragraph{\bf Baselines.} The train-from-scratch (t.f.s) strategy is involved as the baseline. We apply the video dataset (UCF101 \cite{KhurramS_2012}) to pre-train the 3D encoder of the generator on an action recognition task. To transfer the pre-trained weights to the target segmentation task for comparison, the decoder of the generator is randomly initialized. The recently proposed state-of-the-art approaches\footnote{For fair comparison, we pre-train 3D networks on the pretext tasks \cite{Zhou_2019_MICCAI,Zhuang_2019_MICCAI} using experimental datasets, instead of transferring from the publicly available weights \cite{Zhou_2019_MICCAI} pre-trained on external data.} \cite{MedicalNet,Zhou_2019_MICCAI,Zhuang_2019_MICCAI} for the self-supervised and transfer learning of 3D medical data are also involved as benchmark.

\paragraph{\bf Training details.}
Our method is implemented using PyTorch and trained with the Adam~\cite{kingma2014adam} optimizer. The baselines adopt the same training protocol.

\paragraph{Pancreas Rubik's cube.} The pancreas CT is randomly cropped with a size of $128\times128\times128$. The side length of a subcube is set to [7,7,7]. Hence, a Rubik's cube of $18\times18\times18$ is built for each pancreas CT volume. To disarrange the cube, $m=4$ cube layers are randomly rotated along each of the three axes. The Rubik's cube++ recovery task is trained on a GeForce GTX 1080Ti and observed to converge after about 10 hours of training. To preserve the features learned by pre-training, the learning rate is set to 0.0001 while transferring the pre-trained weights to the target task (i.e., pancreas segmentation). The network converges after 250 epochs (about 5 hours) of finetuning with voxel-wise annotations.

\paragraph{Brain Rubik's cube.} The brain MR volume is randomly cropped with a size of $144\times144\times32$. As the number of slice (i.e., $32$) is smaller than the size of axial slices (i.e., $144\times144$), the side length of a subcube is set to [4,4,2], which results in a Rubik's cube with shape of $36\times36\times16$. The other training settings are similar to the prancreas Rubik's cube. As the brain MR scans are multi-modal, we apply the same transformation to each modality and concatenate them as input to the GAN-based network. Due to extremely small size of MRBrainS18, the Rubik's cube++ pre-training is completed in one hour.

\subsection{Ablation Study on Pancreas Segmentation}
The difficulty of Rubik's cube++ is controlled by two parameters: the subcube side length $n$ and the number of rotated layers $m$ in each transformation. However, the grid search of $(n, m)$ is extremely demanding in computation. We suspect that the total number of rotated axial slices $n \times m$ largely determines the difficulty of the pretext task, we fixed one parameter $m=4$ and varying the other (i.e., $n$) for the following experiments.\footnote{An analysis of $m$ can be found in {\itshape arxiv version}.}

To analyze the relationship between the difficulty of Rubik's cube++ and performance improvement to the target task, we construct the pancreas Rubik's cube with different subcube side lengths and evaluate their reconstruction mean squared error (MSE) and pancreas segmentation DSCs with different amount of training data. Table~\ref{tab:parameter} shows the MSE and DSC under different settings of $n$. The larger subcube side length $n$ means more slices are rotated in each transformation, which leads to a lower-order Rubik's cube and makes each cube layer contain more 3D anatomical information. Therefore, rotating such a cube layer may considerably change the inherent structure of organs, which reduces the context information and increases the difficulty for the network to solve the disarranged state.

As shown in Table~\ref{tab:parameter}, the reconstruction error increases from 0.00826 to 0.04762 as the subcube side length increases from 3 to 9. The experimental results are consistent with the finding of existing works \cite{Deng_2010_ECCV,Wei_2019_CVPR}---a harder task often leads to a more robust feature representation. The Rubik's cube++ with $n=7$ achieves the highest DSC for pancreas segmentation with different amounts of training data. There is another finding in our experiments---the pretext is not the harder the better. A performance degradation is observed while $n$ increases to 9, which indicates the excessive destruction of 3D structure may wreck the useful anatomical information for self-supervised learning and consequently decrease the robustness of feature representation.

\begin{table}[!tb]
  \caption{Comparison of DSC (\%) and reconstruction MSE produced by Rubik's cube++ with different $n$ values ($m$ = 4). The listed results are generated via a 4-fold cross validation. (T.f.s.---train-from-scratch)}
  \label{tab:parameter}
  \centering
  \begin{tabular}{c|c|c|c|c|c}
    \hline
                           & \,\,\, T.f.s \,\,\,   & \,\,\, $n = 3$ \,\,\, & \,\,\, $n = 5$ \,\,\, &
    \,\,\, $n = 7$ \,\,\,  & \,\,\, $n = 9$ \,\,\,                                                                       \\
    \hline
    recon. MSE ($1e^{-2}$) & -                     & 0.826                 & 1.609                 & 2.984       & 4.762 \\
    10\%                   & 58.58                 & 47.28                 & 69.72                 & {\bf 73.30} & 68.94 \\
    20\%                   & 70.19                 & 65.00                 & 75.88                 & {\bf 78.10} & 73.34 \\
    50\%                   & 79.68                 & 78.29                 & 81.64                 & {\bf 82.80} & 80.94 \\
    100\%                  & 82.90                 & 82.10                 & 83.72                 & {\bf 84.08} & 83.57 \\
    \hline
  \end{tabular}
\end{table}

\begin{table}[!t]
  \caption{Pancreas segmentation accuracy (DSC \%) of a 4-fold cross validation yielded by 3D U-Nets trained with different strategies. *The result reported in \cite{Wei_2019_CVPR} using V-Net.}
  \label{tab:pancreasS}
  \centering
  \begin{tabular}{l|c|c|c|c}
    \hline
                                            & \,\,\, 10\% \,\,\, & \,\,\, 20\% \,\,\, & \,\,\, 50\% \,\,\, & \,\,\, 100\% \,\,\, \\
    \hline
    Train-from-scratch                      & 58.58              & 70.19              & 79.68              & 82.90               \\
    UCF101 pre-trained                      & 62.21              & 71.90              & 77.14              & 82.76               \\
    Arbitrary puzzles*~\cite{Wei_2019_CVPR} & 70.80              & 76.50              & -                  & 81.68               \\
    MedicalNet~\cite{MedicalNet}            & 64.80              & 71.37              & 77.41              & 80.09               \\
    Rubik's cube~\cite{Zhuang_2019_MICCAI}  & 61.07              & 70.43              & 80.30              & 82.76               \\
    Models genesis \cite{Zhou_2019_MICCAI}  & 63.11              & 70.08              & 79.93              & 83.23               \\
    Rubik's cube++ (Ours)                   & {\bf 73.30}        & {\bf 78.10}        & {\bf 82.80}        & {\bf 84.08}         \\
    \hline
  \end{tabular}
\end{table}

\begin{table}[!t]
  \caption{The mean DSC (\%) of brain tissue segmentation of a leave-one-out cross validation yielded by frameworks trained with different strategies. (T.f.s.---train-from-scratch) The DSC for each class can be found in {\itshape arxiv version}.}
  \label{tab:brain_simple}
  \centering
  \begin{tabular}{l|c|c|c|c|c}
    \hline
                                          & T.f.s & UCF101 pre-trained & Rubik's cube\cite{Zhuang_2019_MICCAI} &
    Models genesis\cite{Zhou_2019_MICCAI} & Ours                                                                                     \\
    \hline
    mean DSC                              & 72.22 & 71.34              & 71.23                                 & 76.19 & {\bf 77.56} \\
    \hline
  \end{tabular}
\end{table}

\subsection{Comparison with State-of-the-art}
\paragraph{\bf Pancreas segmentation.}
The DSCs of 3D U-Nets trained with different strategies via a 4-fold cross validation are presented in Table~\ref{tab:pancreasS}.\footnote{For visual comparison between segmentation results, please refer to {\itshape arxiv version}.} Due to the gap between natural video and medical images, the network finetuned from UCF101 pre-trained weights gains marginal improvement or even degradation with more data used for training, compared to the t.f.s method. Due to the rich information mined from raw data, finetuning from the weights generated by self-supervised learning approaches produces a consistent improvement over the t.f.s strategy. Our method yields the largest increasement to the DSC of pancreas segmentation under all settings of the amount of training data.

\paragraph{Statistical significance.} A t-test validation is conducted on the 4-fold cross validation results (100\% training data) to validate the statistical significance between our Rubik's cube++ and models genesis \cite{Zhou_2019_MICCAI}. A p-value of 3.42\% is obtained, which indicates that the accuracy improvement produced by our approach is statistically significant at the 5\% significance level.

\paragraph{\bf Brain tissue segmentation.}
To further validate the effectiveness of our Rubik's cube++, a leave-one-out experiment is conducted on the MRBrainS18 dataset. The mean DSC of brain tissue segmentation is listed in Table~\ref{tab:brain_simple}. The approaches only pre-training the encoder (i.e., UCF101 and Rubik's cube~\cite{Zhuang_2019_MICCAI}) are observed to deteriorate the mean DSC compared to the train-from-scratch method. In contrast, the context-restoration-based self-supervised learning approaches (models genesis and Rubik's cube++), which simultaneously pre-train the encoder and decoder, generate a significant improvement (i.e., $+3.97\%$ and $+5.34\%$ in mean DSC, respectively) to the brain tissue segmentation task, compared to the train-from-scratch method. The experimental results demonstrate the merit of decoder pre-training for 3D medical image segmentation.

\section{Conclusion}
In this paper, we proposed a context restoration task, i.e., Rubik's cube++, to pre-train 3D neural networks for 3D medical image segmentation. Our Rubik's cube++ adopts a volume-wise transformation for context permutation, which encourages the 3D neural network to better exploit the inherent 3D anatomical information of organs. Our Rubik's cube++ is validated on two publicly available medical datasets to demonstrate its effectiveness, i.e., significantly improving the accuracy of 3D deep learning networks without the use of extra data.

\section*{Acknowledge}
This work is supported by the Key Program of Zhejiang Provincial Natural Science Foundation of China (LZ14F020003), the Natural Science Foundation of China (No. 61702339), the Key Area Research and Development Program of Guangdong Province, China (No. 2018B010111001), National Key Research and Development Project (2018YFC2000702) and Science and Technology Program of Shenzhen, China (No. ZDSYS201802021814180).

%
%
\bibliographystyle{splncs04}
\bibliography{refs}

\newpage

\section*{Appendix}


\begin{table}[!htb]
  \centering
  \caption{Comparison of DSC (\%) yielded by frameworks with different losses on the NIH Pancreas CT dataset. The 4-fold cross validations with different amounts of training data are conducted. Compared to the train-from-scratch (T.f.s.) strategy, Rubik's cube++ with $\mathcal{L}_1$ + $\mathcal{L}_{adv}$ yields the highest improvements.}
  \label{tab:ablation}
  \begin{tabular}{c|c|c|c|c|c}
    \hline
    \,\,\, Training data \,\,\,   & \,\,\, T.f.s. \,\,\,                              & \,\,\, $\mathcal{L}_1$ \,\,\, &
    \,\,\,  $\mathcal{L}_2$ \,\,\,  & \,\,\,  $\mathcal{L}_1$+$\mathcal{L}_{adv}$ \,\,\, &
    \,\,\,  $\mathcal{L}_2$+$\mathcal{L}_{adv}$ \,\,\,                                                                                            \\
    \hline
    20\%                        & 70.19                                           & 72.14                       & 72.28 & {\bf 78.10} & 73.75 \\
    100\%                       & 82.90                                           & 83.22                       & 83.95 & {\bf 84.08} & 83.85 \\
    \hline
  \end{tabular}
\end{table}

\begin{figure*}[!htb]
    \centering
    \includegraphics[width=\textwidth]{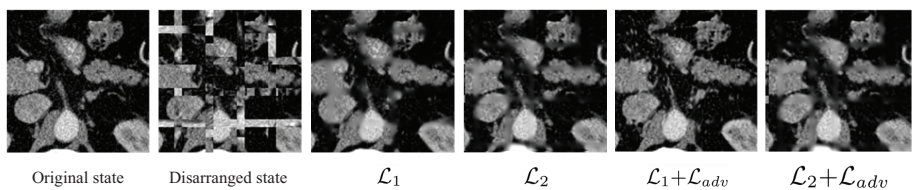}
    \caption{Restoration results produced by our framework with different losses on the NIH Pancreas CT dataset. Compared to $\mathcal{L}_{1}$, the $\mathcal{L}_{2}$ loss is inclined to have blurry solutions, which may lose the boundary information of organs.} \label{fig:reconstruction}
\end{figure*}


\begin{table}[!htb]
  \centering
  \caption{Comparison of DSC (\%) produced by Rubik's cube++ with different $m$ values ($n$ = 7) on the NIH Pancreas CT dataset. A 4-fold cross validation with 20\% training data is conducted. The increase of $m$ means more cube layers are rotated in each transformation, resulting in a harder disarranged Rubik's cube for 3D neural networks to resolve. (T.f.s.---train-from-scratch)}\label{tab:parameter_m}
  \begin{tabular}{c|c|c|c|c}
    \hline
    {}   & T.f.s. & $m = 3$ & $m = 4$     & $m = 5$ \\\hline\hline
    \,\,\, 20\% \,\,\,  & \,\,\, 70.19 \,\,\,  & \,\,\, 68.08 \,\,\,   & \,\,\, {\bf 78.10} \,\,\, & \,\,\, 73.46 \,\,\,   \\\hline
  \end{tabular}
\end{table}


\begin{figure*}[!htb]
  \footnotesize
  \begin{minipage}[t]{0.115\textwidth}
    \centering
    \includegraphics[width=\textwidth]{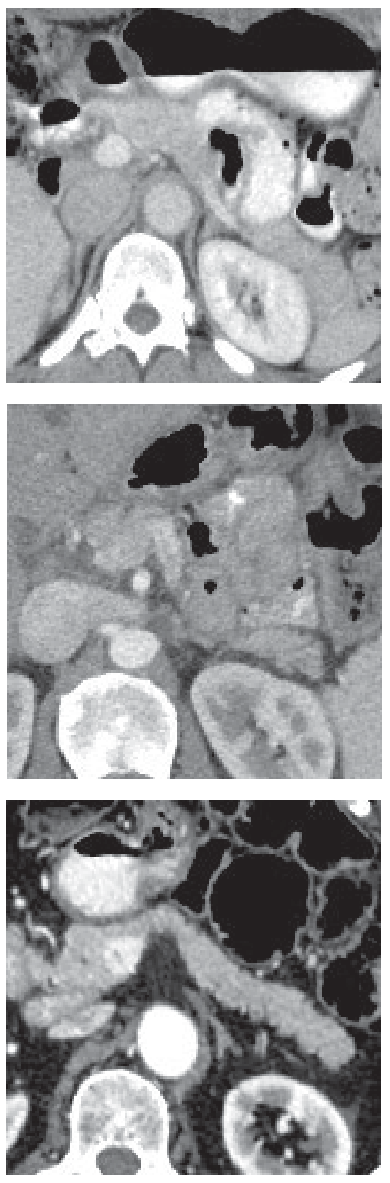}
    {(a) Original image}
  \end{minipage}
  \begin{minipage}[t]{0.115\textwidth}
    \centering
    \includegraphics[width=\textwidth]{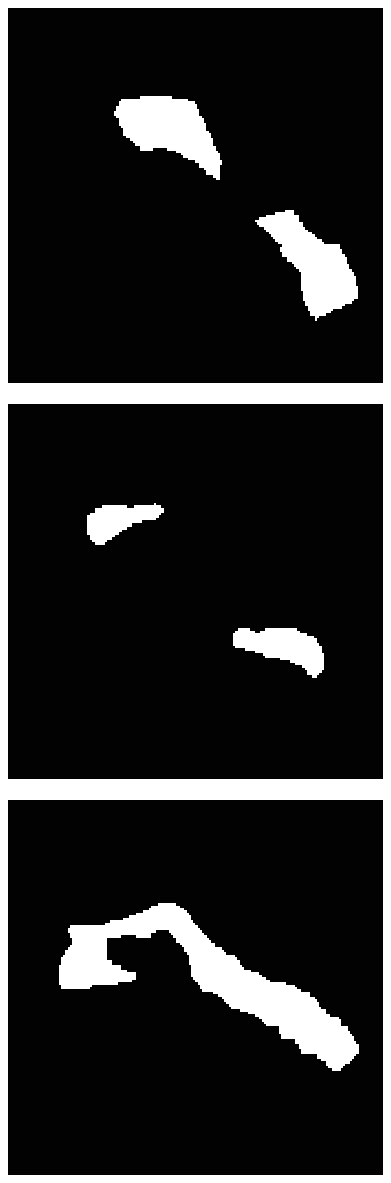}
    {(b) Ground truth}
  \end{minipage}
  \begin{minipage}[t]{0.115\textwidth}
    \centering
    \includegraphics[width=\textwidth]{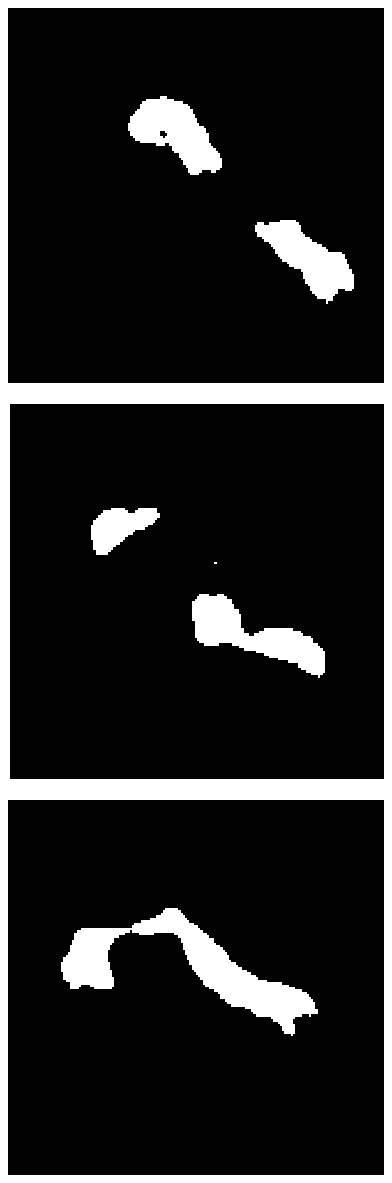}
    {(c) T. f. s.}
  \end{minipage}
  \begin{minipage}[t]{0.115\textwidth}
    \centering
    \includegraphics[width=\textwidth]{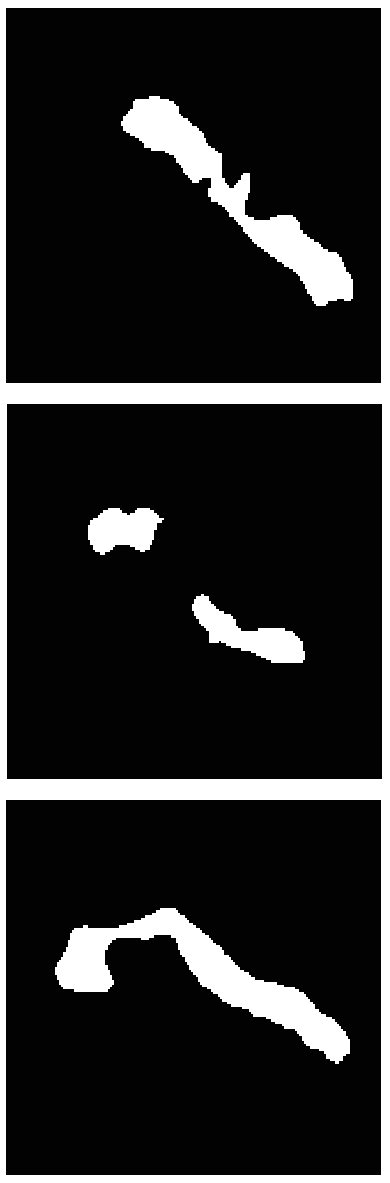}
    {(d) UCF101}
  \end{minipage}
  \begin{minipage}[t]{0.115\textwidth}
    \centering
    \includegraphics[width=\textwidth]{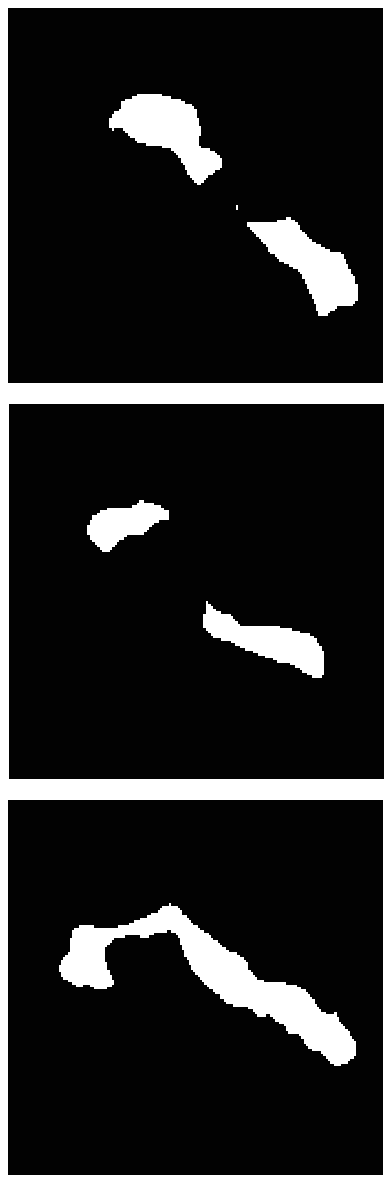}
    {(e) Rubik's cube \cite{Zhuang_2019_MICCAI}}
  \end{minipage}
  \begin{minipage}[t]{0.115\textwidth}
    \centering
    \includegraphics[width=\textwidth]{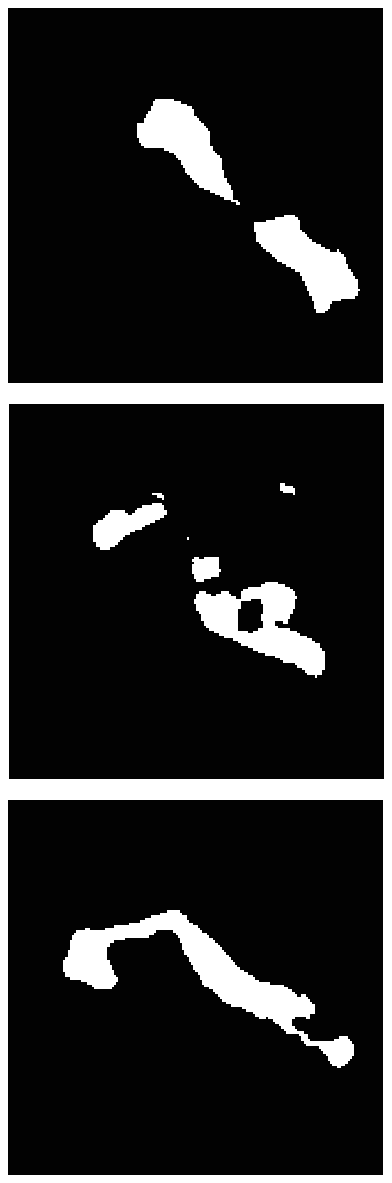}
    {(f) Models genesis \cite{Zhou_2019_MICCAI}}
  \end{minipage}
  \begin{minipage}[t]{0.115\textwidth}
    \centering
    \includegraphics[width=\textwidth]{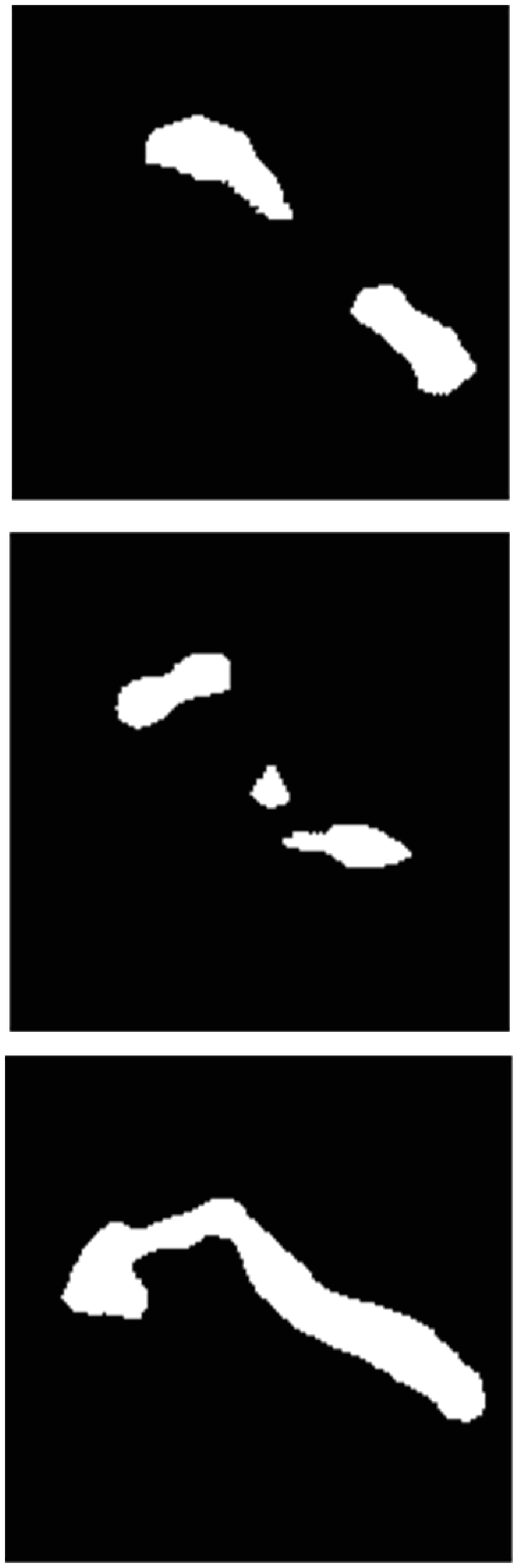}
    {(g) MedicalNet \cite{MedicalNet}}
  \end{minipage}
  \begin{minipage}[t]{0.115\textwidth}
    \centering
    \includegraphics[width=\textwidth]{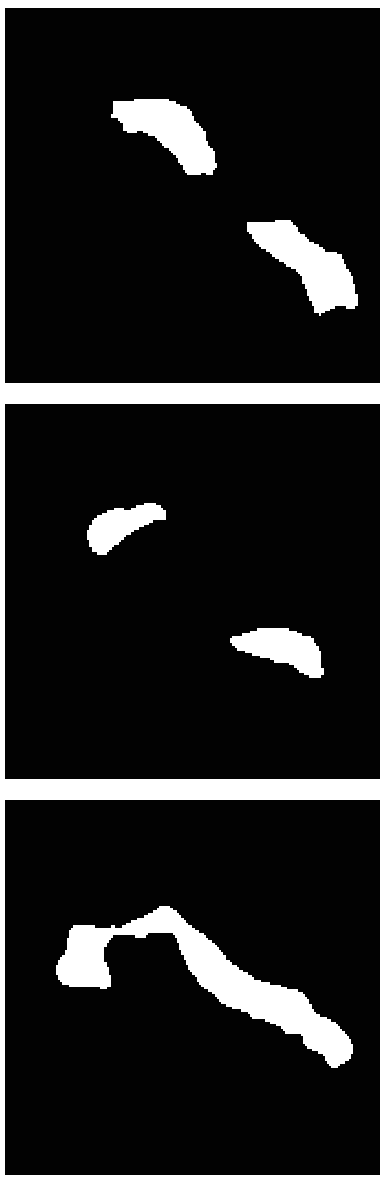}
    {(h) Rubik's cube++ (ours)}
  \end{minipage}
  \caption{Segmentation results of NIH Pancreas-CT (100\% training data) generated by 3D U-Nets with different training strategies. (T. f. s. refers to train-from-scratch)} \label{fig:segmentation_res_100}
\end{figure*}

\begin{figure*}[!htb]
  \footnotesize
  \begin{minipage}[t]{0.134\textwidth}
    \centering
    \includegraphics[width=\textwidth]{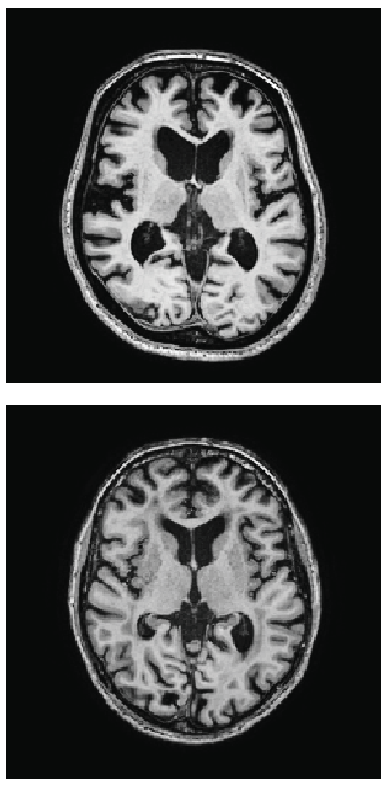}
    {(a) Original image}
  \end{minipage}
  \begin{minipage}[t]{0.134\textwidth}
    \centering
    \includegraphics[width=\textwidth]{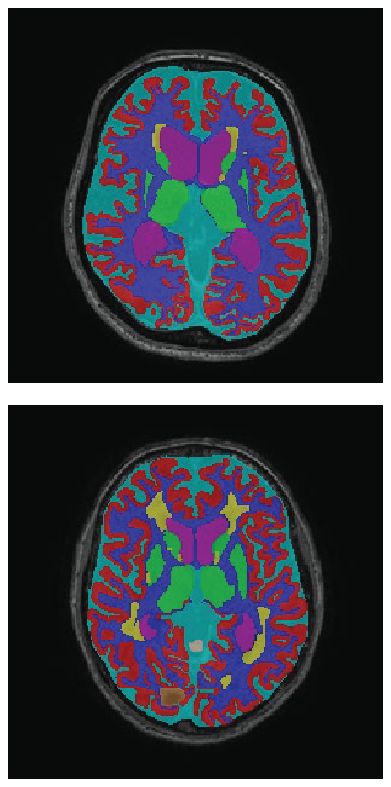}
    {(b) Ground truth}
  \end{minipage}
  \begin{minipage}[t]{0.134\textwidth}
    \centering
    \includegraphics[width=\textwidth]{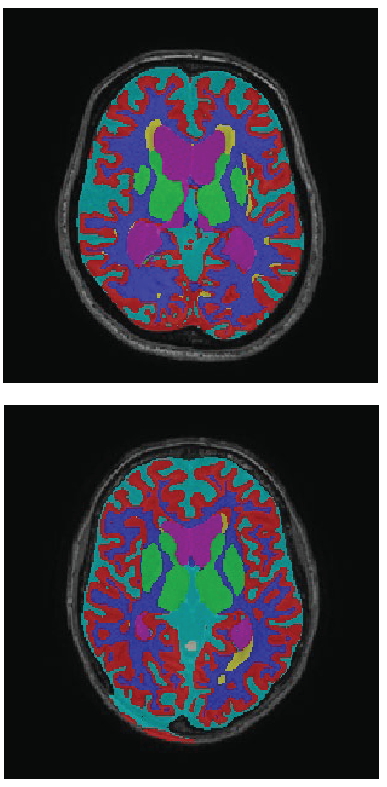}
    {(c) T. f. s.}
  \end{minipage}
  \begin{minipage}[t]{0.134\textwidth}
    \centering
    \includegraphics[width=\textwidth]{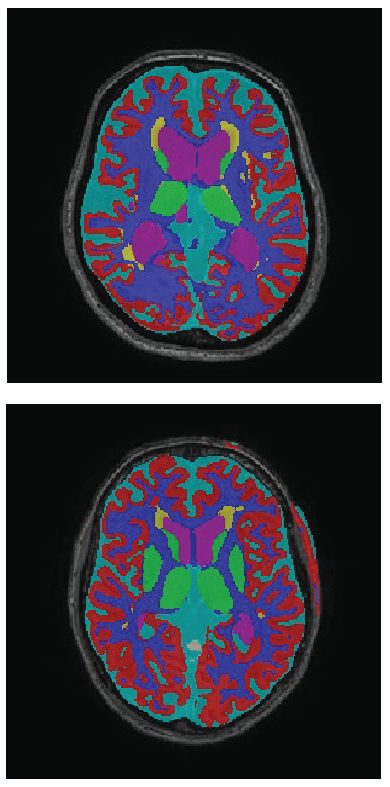}
    {(d) UCF101}
  \end{minipage}
  \begin{minipage}[t]{0.134\textwidth}
    \centering
    \includegraphics[width=\textwidth]{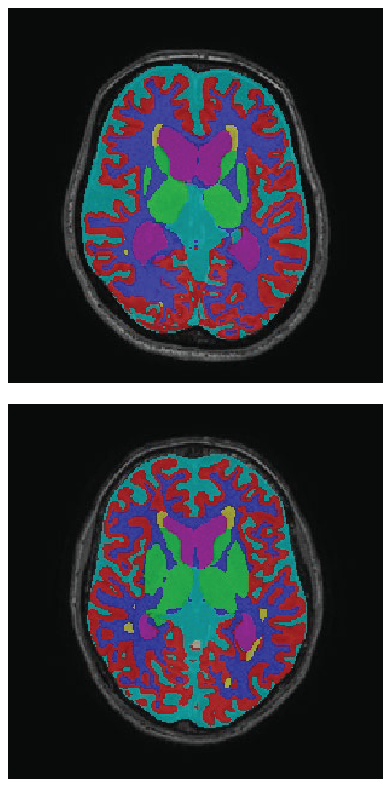}
    {(e) Rubik's cube \cite{Zhuang_2019_MICCAI}}
  \end{minipage}
  \begin{minipage}[t]{0.134\textwidth}
    \centering
    \includegraphics[width=\textwidth]{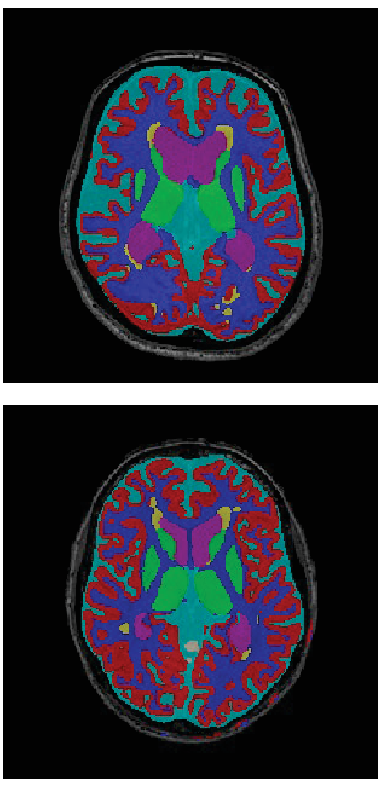}
    {(f) Models genesis \cite{Zhou_2019_MICCAI}}
  \end{minipage}
  \begin{minipage}[t]{0.134\textwidth}
    \centering
    \includegraphics[width=\textwidth]{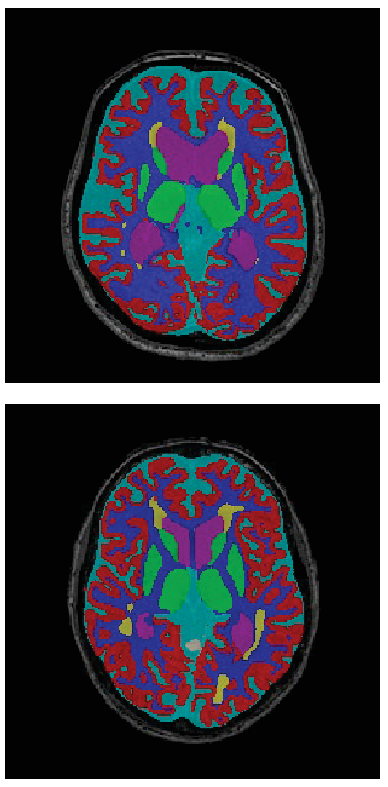}
    {(g) Rubik's cube++ (ours)}
  \end{minipage}
  \caption{Segmentation results of MRBrainS18 generated by 3D U-Nets trained with different strategies. (T. f. s. refers to train-from-scratch)} \label{fig:segmentation_res_70148}
\end{figure*}

\begin{table*}[!htb]
  \centering
  \caption{Brain tissue segmentation accuracy of leave-one-out cross validation yielded by 3D U-Nets trained with different strategies on the MRBrainS18 dataset. (C. gray matter---Cortical gray matter, W. m. lesions---White matter lesions)} \label{tab:fully_sup_structures}
  \setlength{\tabcolsep}{0.2mm}{
    \begin{tabular}{p{3.3cm}|P{0.9cm}P{0.9cm}P{0.9cm}P{0.9cm}P{0.9cm}P{0.9cm}P{0.9cm}P{0.9cm}P{0.9cm}P{0.9cm}}
                                             & \rotatebox{66}{C. gray matter} & \rotatebox{66}{Basal ganglia} & \rotatebox{66}{White matter} & \rotatebox{66}{W. m. lesions} & \rotatebox{66}{CFS} & \rotatebox{66}{Ventricles} & \rotatebox{66}{Cerebellum} & \rotatebox{66}{Barin stem} & \rotatebox{66}{Mean DSC} \\\hline\hline
      Train-from-scratch                     & 78.03                          & 72.57                         & 79.91                        & 35.42                         & 70.90               & 90.86                      & 85.69                      & 70.15                      & 72.22                    \\
      UCF101 pre-trained                     & 77.49                          & 73.79                         & 79.42                        & 30.98                         & 76.49               & 90.39                      & 83.26                      & 65.05                      & 71.34                    \\
      Rubik's cube \cite{Zhuang_2019_MICCAI} & 77.31                          & 73.09                         & 78.60                        & 34.64                         & 71.96               & 88.97                      & 83.52                      & 67.82                      & 71.23                    \\
      Models genesis \cite{Zhou_2019_MICCAI} & 80.35                          & {\bf 80.13}                   & 82.26                        & {\bf 44.95}                   & 77.34               & 92.12                      & 87.63                      & 68.90                      & 76.19                    \\
      Rubik's cube++ (ours)                  & {\bf 81.42}                    & 79.63                         & {\bf 83.20}                  & 44.81                         & {\bf 78.69}         & {\bf 92.46}                & {\bf 88.32}                & {\bf 75.81}                & {\bf 77.56}              \\
    \end{tabular}}
\end{table*}

\end{document}